# New Basic Form of the Semiclassical Quantization Condition

N.N.Trunov[*]

*D.I.Mendeleyev Institute for Metrology*

*Russia, St.Peterburg. 190005 Moskovsky pr. 19*

(Dated: December 10, 2008)

**Abstract :** A certain modification of the semiclassical quantization condition based on the summarization of the known power expansion series in the squared Planck constant is proposed. Corresponding deviation from exact spectra arises only together with the deviation of the studied potential from the supersymmetryc one so that on the base of the presented method a new kind of the perturbation theory not connected with the Planck constant may be developed.

## 1. Introduction

Suppose we determine the energy levels $\varepsilon_n$ in a potential well $V(x)$. In general case we can write the corresponding equation in the following form:

$$\int \sqrt{2m(\varepsilon - V)}\, dx = \pi \hbar (n + \tfrac{1}{2}) + \Delta \qquad (1)$$

where $n = 0, 1, 2...$ and the integral is taken between turning points. Here $\Delta$ is an unknown yet function of $\varepsilon$, and also a functional of $V$ which ensures the exact spectrum.

---

[*] Electronic adress: trunov@vniim.ru

Neglecting $\Delta$, i.e. putting $\Delta \equiv 0$ we get the usual basic condition for the semiclassical quantization. Since it is not exact for almost all practically used potentials many attempts were undertaken to improve such condition. They may be divided into two groups. On the one hand there were proposed [1] various forms of $\Delta$ adapted for special subclasses of potentials ad hoc with respect to asymptotics of the exact solutions or other arguments outside the semiclassical approach itself. On the other hand a power series expansion of $\Delta$ in $\hbar^2$ was found [2]:

$$\Delta = \sum_{k=1}^{\infty} \Delta_k,$$
$$\Delta_k = f_k(\varepsilon)\hbar^{2k}.$$

(2)

The higher order $\Delta_k$ we calculate, the more cumbersome they are and the higher orders of derivatives of $V$ they include. That's why one use $\Delta$ very seldom and even in this case only as

$$\Delta_1 = \frac{\hbar^2}{24\sqrt{2m}} \frac{\partial^2}{\partial \varepsilon^2} \int \frac{dx}{\sqrt{(\varepsilon - V)}} \left(\frac{\partial V}{\partial x}\right)^2$$

(3)

Thus the equation (1) with $\Delta \equiv 0$ remains the basic form discussed in all handbooks.

In the present paper we propose another simple enough basic equation of the semiclassical quantization with $\Delta$ as some definite function of $\Delta_1$. This condition ensures exact spectra for all potentials listed in the famous paper [3], for all reference potentials mentioned in handbooks treating the semiclassical approach, and has many other advantages discussed below.

## 2. Calculations of $\Delta$

A very wide class of usually used model potentials can be written in the following form:

$$V(x) = A^2 s^2 + Bs + C$$

$$\sigma \equiv \frac{ds}{dx} = a_2 s^2 + a_1 s + a_0 > 0 \qquad (4)$$

with an auxiliary function $s(x)$. For example, the oscillator potential

$$V(x) = \frac{A^2 x^2}{2} \qquad (5)$$

$$s = \frac{x}{\sqrt{2}}, \quad \sigma = \frac{1}{\sqrt{2}}$$

the hyperbolic potential well

$$V(x) = -A^2 ch^{-2} x = A^2 s^2 - A^2 \qquad (6)$$

$$s = thx, \quad \sigma = 1 - s^2$$

and the Coulomb potential with the centrifugal one

$$V = A^2 s^2 + Zs \qquad (7)$$

$$s = -\frac{1}{x}; \quad \sigma = s^2$$

The simplest way to determine $\Delta$ as some function of $\Delta_1$ is to compare known exact spectra for potentials (4) with (1). For $V$ (6) we write the result [1] in a suitable form

$$\sqrt{2m}\left(A - \sqrt{\varepsilon}\right)\pi = \pi\left(n + \frac{1}{2}\right) - A\pi\sqrt{2m}\left[\sqrt{1 + \frac{\hbar^2}{8A^2 m}} - 1\right]. \qquad (8)$$

The left side of (8) coincides with the usual semiclassical condition, so that the second term in the right side is equal to $\Delta$:

$$\Delta = -\frac{\pi\hbar^2 \sqrt{2}}{8A\sqrt{m}\left[\sqrt{1 + \frac{\hbar^2}{8A}} + 1\right]}. \qquad (9)$$

Here we used the identity:

$$\sqrt{1+x} - 1 = \frac{x}{\sqrt{1+x}+1}.$$

Obviously the first term of the power expansion of $\Delta$ in $\hbar^2$ is equal to $\Delta_1$:

$$\Delta_1 = -\frac{\pi \hbar^2}{8\sqrt{2mA}}. \tag{10}$$

Now expressing $\hbar^2/A$ through $\Delta_1$ we get:

$$\Delta = \frac{2\Delta_1}{1+\sqrt{1+\frac{16}{\pi^2 \hbar^2}\Delta_1^2}}. \tag{11}$$

Using known results [1] it is easy to prove that (11) is valid for all potentials (4) at any values $B$, $C$, $a_1$, $a_0$ (and not only for zero ones).

Note that (10) is certainly identical to

$$\Delta_1 = \frac{\pi \hbar^2 a_2}{8A\sqrt{2m}}, \tag{12}$$

obtained from (3) for all potentials (6).

Thus the quantization condition (1) with $\Delta$ (11), (10) is exact within the class of potentials (4). For these potentials two expressions (3) and (10) lead obviously to the same value of $\Delta_1$, but it is not the fact for any potentials outside the class (14). Values of $a_2$ and $\Delta_1$ do not change if we transform $V$ (4) to the standard form $V = A^2 s^2$ by means of linear shifts.

## 3. Properties of the improved condition

We propose as an improved basic form of the semiclassical quantization condition the expression (1) with $\Delta$ (11), where the general form (3) of $\Delta_1$ is substituted in. Introducing a new parameter

$$\beta^2 = \frac{\hbar^2}{2m}$$

we rewrite (1), (11), (3) in the following explicit form:

$$\frac{1}{\pi \beta}\int \sqrt{\varepsilon - V}\,dx = n + \frac{1}{2} + \delta \tag{13}$$

$$\delta = \frac{2\delta_1}{1+\sqrt{1+16\delta_1^2}} \quad ; \qquad \delta_1 = \frac{\beta}{24\pi}\frac{\partial^2}{\partial\varepsilon^2}\int\frac{dx}{\sqrt{\varepsilon-V}}\left(\frac{\partial V}{\partial x}\right)^2$$

The dimensionless shift $\delta$ clearly indicates the new correction to the usual quantization condition. In the limiting case $|\delta_1|\to\infty$ which may occur at the very small amplitude $A$, see (6),

$$\delta = \frac{1}{2}\operatorname{sgn}\delta_1, \qquad (14)$$

so that instead usual $n+\tfrac{1}{2}$ we get $n+1$ or $n$. Our quantization condition (13) certainly remains correct for potentials (4) also in this limiting case. For example, in (7) $A^2 = \hbar^2 l(l+1)$ if $l$ is the orbital quantum number. Then (13) ensures the exact spectrum even for $l=0$ i.e. $A\to 0$.

Moreover we can easily demonstrate the correctness of (14) for the whole important class of potentials outside (4), namely, for the potential wells with coinciding asymptotic values:

$$\begin{aligned} V(x) &\to V_0 \\ x &\to \pm\infty; \end{aligned} \qquad (15)$$

here we can put $V(0)=0$. It is known for all such potentials that the lower level with $n=0$ exists at any amplitude $V_0$, even if $V_0\to 0$ (i.e. $A^2\to 0$ in (6)). Since the left side of (13) also approaches to zero when $V_0\to 0$, in the right side (13) at $n=0$ we must become $\delta\to -1/2$. But this fact really takes place at large $\delta_1$, i.e. small amplitude $V_0$ independently on the explicit form $V$ satisfying (15).

In the opposite limiting case of small $|\delta_1|$ we have $\delta$ practically coinciding with $\delta_1$ so that our condition (13) is again valid for all potentials. Note that $\delta$ changes its order as some power of $\beta$ when the value of $\delta_1$ changes: $\delta = const\,\beta$ at small $|\delta_1|$ and does not depend on $\beta$ at large $|\delta_1|$. Respectively our condition as a whole is not a power expansion in $\beta$.

Practical correctness of our condition in both limiting cases $|\delta_1| \to 0$, $|\delta_1| \to \infty$ for a very wide set of potentials makes to expect that this new condition will be also effective for intermediate values of $\delta_1$.

As it was already said, our condition is exact for all the potentials listed in [3] and suitable for the factorization method. These potentials may also be treated as supersymmetric ones with an additive parameter [4, 5].

All or almost all interesting potentials are situated in some vicinity of those belonging to (4). Thus a new type of the perturbation theory may be developed with some small parameter describing a deviation from properties of the supersymmetry or the "factorizability". Such small parameter may include directly or indirectly $\gamma = d\delta_1/d\varepsilon$, since $\gamma \equiv 0$ for all potentials (4). It should be stressed once more that the hard algebraic construction (4) built in the usual semiclassical approach allows to construct the perturbation theory independent on $\beta$.

Meanwhile we can expect that even the zero order in $\gamma$ approximation (13) must give good results; another, more cumbersome variant of this approximation was presented in [5].